\documentstyle[sprocl]{article}

\input{psfig}

\def\Journal#1#2#3#4{{#1} {\bf #2}, #3 (#4)}

\def\NPA{{\em Nucl. Phys.} A} 
\def\PLB{{\em Phys. Lett.}  B}
\def\PRL{\em Phys. Rev. Lett.}
\def\PRD{{\em Phys. Rev.} D}

\def\PAN{\em Phys. At. Nuclei }
\def\EPJ{{\em Eur. Phys. J.} C }

\def\be{\begin{equation}}
\def\ee{\end{equation}}
\def\bea{\begin{eqnarray}}
\def\eea{\end{eqnarray}}

\begin{document}

\title{Light-front quark model predictions of meson elastic and
       transition form factors }

\author{Chueng-Ryong Ji and Ho-Meoyng Choi}

\address{ Department of Physics, North Carolina State University,
Raleigh,\\ NC 27695-8202, USA\\E-mail: ji@ncsu.edu} 

\maketitle\abstracts{ 
We investigate the electroweak form factors and
semileptonic decay rates of mesons using the constituent
quark model based on the light-front degrees of freedom.
Our results demonstrate the broader applicability of light-front 
approach including the timelike region of exclusive processes.}

\section{Introduction}

One of the distinctive advantages in the light-front approach is the
well-establi-

\noindent
shed formulation of various form factor calculations 
using the well-known Drell-Yan-West ($q^{+}=0$) frame~\cite{LB}.
In $q^{+}=0$ frame, only parton-number-conserving Fock
state (valence) contribution is needed when the ``good" components of the
current, $J^{+}$ and $J_{\perp}=(J_{x},J_{y})$, are used~\cite{CJ2}.
For example, only the valence diagram shown in Fig. 1(a) is used in the
light-front quark model (LFQM) analysis of spacelike meson form factors.
Successful LFQM description of various hadron form
factors can be found in the literatures~\cite{CJC,CJ4,CJ1}.

However, the timelike ($q^{2}>0$) form factor analysis in the LFQM 
has been hindered by the fact that $q^{+}=0$ frame is defined
only in the spacelike region ($q^{2}=q^{+}q^{-}-q^{2}_{\perp}<0$).
While the $q^{+}\neq0$ frame can be used in principle to compute the
timelike form factors, it is inevitable (if $q^{+}\neq 0$) to encounter the
nonvalence diagram arising from the quark-antiquark pair creation (so called
``Z-graph"). For example, the nonvalence diagram in the case of
semileptonic meson decays is shown in Fig. 1(b).
The main source of the difficulty, however, in calculating the nonvalence
diagram (see Fig. 1(b)) is the lack of information on the black blob
which should contrast with the white blob representing the usual light-front
valence wave function. In fact, we noticed~\cite{CJ2} that
the omission of nonvalence contribution
leads to a large deviation from the full results.
The timelike form factors associated with the hadron pair productions in
$e^{+}e^{-}$ annihilations also involve the nonvalence contributions.
Therefore, it would be very useful to avoid encountering the nonvalence
diagram and still be able to generate the results of timelike form factors.
\begin{figure}[t]
\psfig{figure=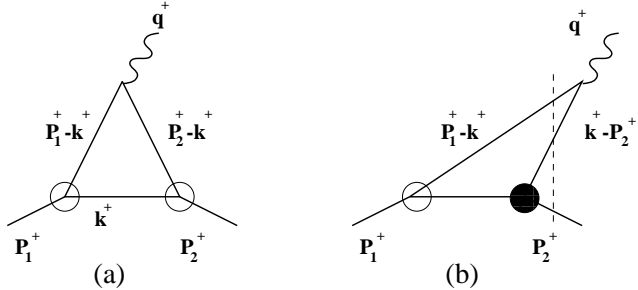,height=1.5in}
\caption{The LFQM description of a electroweak
meson form factor: (a) the usual light-front valence diagram and (b) the
nonvalence(pair-creation) diagram. The vertical dashed line in (b)
indicates the energy-denominator for the nonvalence contributions.
While the white blob represents the usual light-front
valence wave function, the modeling of black blob has not yet been made.
\label{fig:triangle}}
\end{figure}

In this paper, we start from the heuristic model calculations 
to obtain the exact result of the timelike form factor and then  discuss
our light-front constituent quark model predictions for the meson
electroweak form factors. 
Especially, we focus on the calculations of observables in timelike
$|Q^{2}|$ region overcoming the diffculties associated with the
quark-antiquark pair creation in the light-front approach. 
As an explicit example of application to timelike
$|Q^{2}|$ region, we present our recent analysis of semileptonic weak decay
processes. In Section 2, we present the heuristic model calculations to
analytically continue the form factors in the spacelike region to those 
in the timelike region. In Section 3, we present
our model calculations of semileptonic pseudoscalar meson decay processes
as an example of application to the timelike $|Q^{2}|$ region. Conclusions
and discussions follow in Section 4.

\section{Heuristic Model Calculations}
Using the solution of covariant Bethe-Salpeter equation in the ladder
approximation with a relativistic version of the contact 
interaction~\cite{SM}, we can learn some lessons  
on the analytic continuation from the spacelike region to 
the timelike region~\cite{CJ8}. The covariant model wave function 
is a product of two free single particle propagators, the overall 
momentum-conservation Dirac delta and a constant vertex function. 
Thus, all our form factor calculations in this section are nothing
but various ways of evaluating the Feynman perturbation theory triangle
diagram in scalar field theory~\cite{CJ8}. 

Using this model, we show an explicit example of generating the exact result
of the timelike form factor without encountering the nonvalence diagram.
This can be done by the analytic continuation from the spacelike form
factor calculated in the Drell-Yan-West ($q^{+}=0$) frame to the timelike
region. To explicitly show it, we calculated~\cite{CJ8}:
(A) the timelike process of $\gamma^{*}\to M + \bar{M}$
transition in $q^{+}\neq 0$ ($q^{2}>0$) frame,
(B) the spacelike process of $M\to\gamma^{*} +M$ in $q^{+}\neq0$
($q^{2}<0$) frame, and
(C) the spacelike process of $M\to\gamma^{*} +M$ in $q^{+}=0$ frame.
The analytic continuation from $q^{2}<0$ to $q^{2}>0$ allows us to 
show that the result in (C), which is obtained without encountering the
nonvalence contributions at all, exactly reproduces the result in (A).
In fact, all three results (A), (B), and (C) coincide with each other in
the entire $q^{2}$ range.
We also confirm that our results are consistent with the dispersion
relations~\cite{Drell}. We consider here only for the equal
quark/antiquark mass case such as the pion. However, the unequal
mass cases such as $K$ and $D$ were also shown in Ref.[7].

For our numerical analysis of $\pi$ meson form factor,
we use the following constituent quark and antiquark masses: 
$m_{u}=m_{d}=0.25$ GeV.
Since our numerical results of the EM form factors obtained from 
(A), (B), and (C) turn out to be exactly same with each other for the entire
$q^{2}$ region, each line depicted in Figs. 2 and 3 represents all three
results.

As a consistency check, we also compare
our numerical results of the form factor $F(q^{2})= {\rm Re}\;
F(q^{2}) + i\;{\rm Im}\; F(q^{2})$ with the dispersion relations given by
\begin{eqnarray}
{\rm Re}\;F(q^{2})&=&\frac{1}{\pi}P\int^{\infty}_{-\infty}
\frac{{\rm Im}\;F(q'^{2})}{q'^{2}-q^{2}}dq'^{2},\\
{\rm Im}\;F(q^{2})&=&-\frac{1}{\pi}P\int^{\infty}_{-\infty}
\frac{{\rm Re}\;F(q'^{2})}{q'^{2}-q^{2}}dq'^{2},
\end{eqnarray}
where $P$ indicates the Cauchy principal value.

\begin{figure}[t]
\psfig{figure=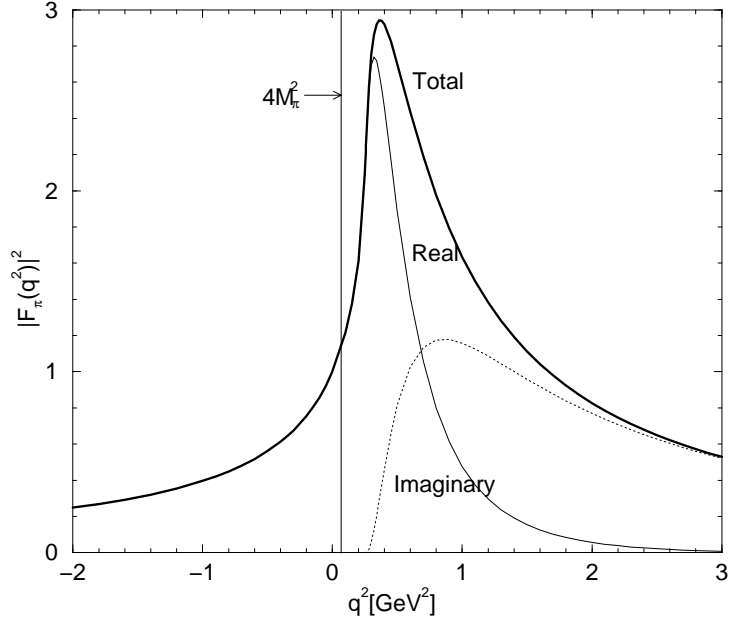,height=3.8in}
\caption{
The electromagnetic form fator of the pion in $(3+1)$ dimensional
scalar field theory for $-2\leq q^{2}\leq 3$ GeV$^{2}$. The total,
real, and imaginary parts of $|F_{\pi}(q^{2})|^{2}$ are represented
by thick solid, solid, and dotted lines, respectively. 
\label{fig:pion1}}
\end{figure}
\begin{figure}[t]
\psfig{figure=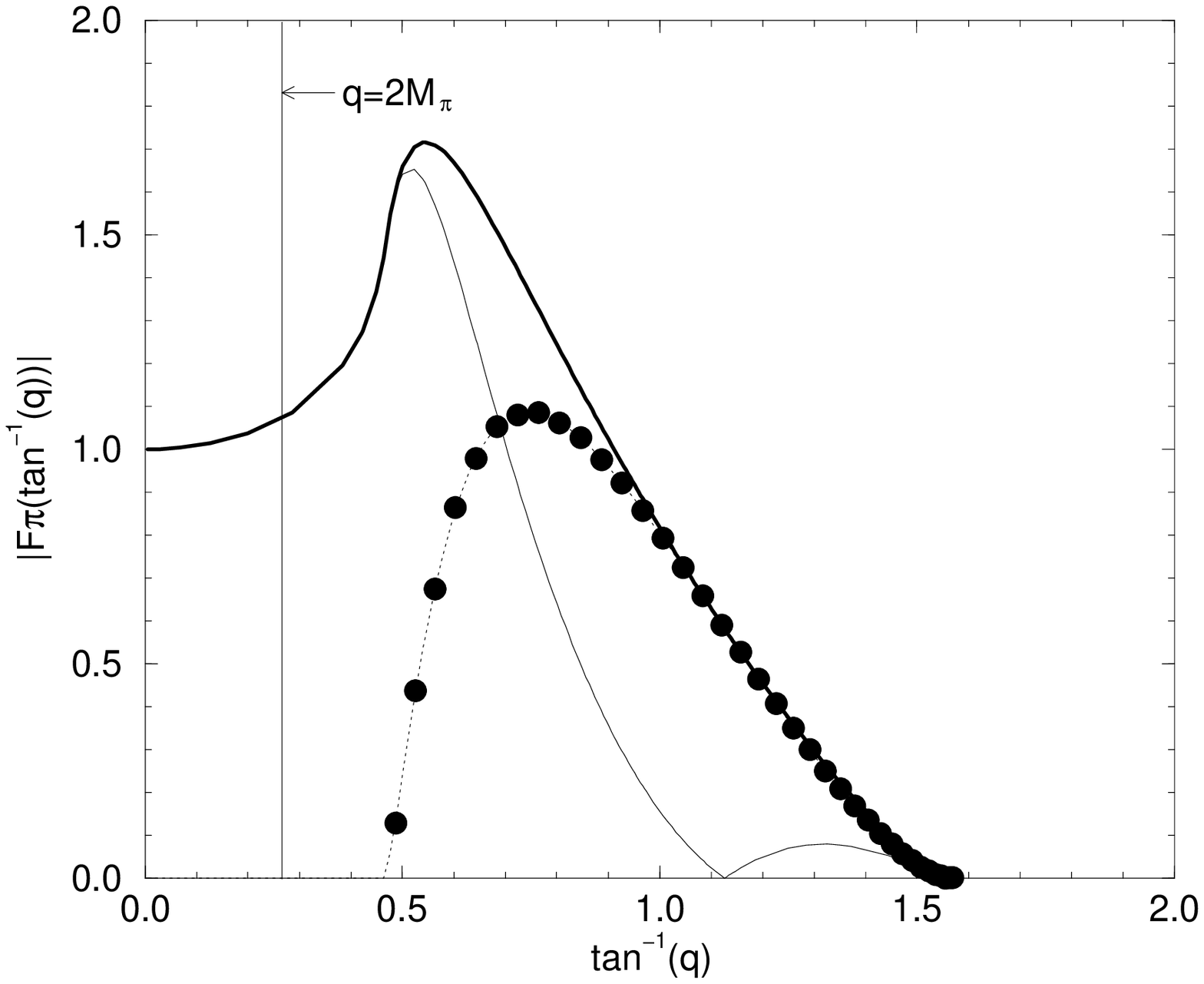,height=3.8in}
\caption{ The electromagnetic form fator of the pion in $(3+1)$ dimensional
scalar field theory for the entire timelike region compared to the
dispersion relations (data of black dots) given by Eq. (2).
The same line code as in Fig. 2 is used. \label{fig:pion2}}
\end{figure}
In Fig. 2, we show the EM form factor of the pion for
$-2\;{\rm GeV}^{2}\leq q^{2}\leq 3\;{\rm GeV}^{2}$.
The imaginary part (the dotted line)
of the form factor starts at $q^{2}_{\rm min}=4m^{2}_{u(d)}=0.25$ GeV.
It is interesting to note that
the square of the total form factor $|F_{\pi}(q^{2})|^{2}$ (thick solid
line) produces a $\rho$ meson-type peak near $q^{2}\sim M^{2}_{\rho}$.
However, it is not yet clear if this model indeed reproduces all the
features of the vector meson dominance (VMD) phenomena. Even though
the generated position of peak is consistent with VMD, the final state
interaction is not included in this simple model calculation.
We believe that much more complex mechanisms may be necessary to
reproduce the realistic VMD phenomena. More detailed analysis along this
line is under consideration. Nevertheless, it is remarkable that this
simple model is capable of generating the peak and the position
of peak is quite consistent with the VMD.

In Fig. 3, we show the timelike form factor of the pion for the entire
$q^{2}>0$ region and compare the imaginary part of our direct
calculations (dotted line) obtained from (A), (B), and (C) with the result
(data of black dots) obtained from the dispersion relations given by
Eq. (2). Our direct calculation is in an excellent agreement
with the solution of the dispersion relations. Our result for the real
part are also confirmed to be in complete agreement with the dispersion
relations. For high $q^{2}$ region, the imaginary part of the form 
factor is dominant over the real part (thin solid line).

In Figs. 2 and 3, it is astonishing that the
numerical result of (C)
obtained from $q^{+}=0$ frame without encountering the nonvalence diagram
coincides exactly with the numerical results of (A) and (B)
obtained from $q^{+}\neq 0$ frame.

\section{ Semileptonic Pseudoscalar Meson Decays in LFQM}
The key idea in our LFQM~\cite{CJ1} for mesons is to treat
the radial wave function as a trial function for the variational
principle to the QCD-motivated Hamiltonian saturating
the Fock state expansion by the \underline {constituent} quark and
antiquark. The spin-orbit wave function is uniquely determined by the 
Melosh transformation.  We take the QCD-motivated effective Hamiltonian 
as the well-known linear plus Coulomb interaction given by
\begin{eqnarray}
H_{q\bar{q}}= H_{0} + V_{q\bar{q}}
=\sqrt{m_{q}^{2}+k^{2}} + \sqrt{m_{\bar{q}}^{2}+k^{2}}+ V_{q\bar{q}},
\end{eqnarray}
where
\begin{eqnarray}
V_{q\bar{q}}= V_{0} + V_{\rm hyp}
= a + br - \frac{4\kappa}{3r}
+ \frac{2\vec{S}_{q}\cdot\vec{S}_{\bar{q}}}
{3m_{q}m_{\bar{q}}}\nabla^{2}V_{\rm Coul}.
\end{eqnarray}
We take the Gaussian radial wave function 
$\phi(k^{2})=N\exp(-k^{2}/2\beta^{2})$
as our trial wave function to minimize the central Hamiltonian~\cite{CJ1}. 
Since the string tension $b$=0.18 GeV$^{2}$ and the constituent $u$ and 
$d$ quark masses $m_{u}$=$m_{d}$=0.22 GeV are rather well known from 
other quark model analyses commensurate with Regge 
phenomenology~\cite{Isgur}, we take them as our input parameters.
The model parameters of $a,\kappa$, and $\beta_{u\bar{d}}$ 
are determined by the variational principle using the masses
of $\rho$ and $\pi$~\cite{CJ1,CJ9}. It is very important to note that
all other model parameters such as $m_{c}$, $m_{b}$,
$\beta_{uc}$, $\beta_{ub}$, etc. are then uniquely determined by 
our variational principle as shown in Ref.[10]. 

The semileptonic decays of a pseudoscalar meson $Q_{1}\bar{q}$ into
another pseudoscalar meson $Q_{2}\bar{q}$ are governed by the weak vector
current as follows
\begin{eqnarray}
\langle P_{2}|\bar{Q_{2}}\gamma^{\mu}Q_{1}|P_{1}\rangle&=&
f_{+}(q^{2})(P_{1}+P_{2})^{\mu} + f_{-}(q^{2})(P_{1}-P_{2})^{\mu},
\end{eqnarray}
where $q^{\mu}=(P_{1}-P_{2})^{\mu}$ is the four-momentum
transfer to the leptons.
In the LFQM calculations presented in Ref.[11], the 
$q^{+}\neq 0$ frame has been used to calculate the weak decays
in the timelike region $m^{2}_{l}\leq q^{2}\leq (M_{1}-M_{2})^{2}$, with
$M_{1[2]}$ and $m_{l}$ being the initial[final] meson mass and the
lepton($l$) mass, respectively.
However, when the $q^{+}\neq 0$ frame is used, the inclusion of the
nonvalence contributions arising from quark-antiquark pair
creation(``Z-graph") is inevitable and this inclusion may be
very important for heavy-to-light and light-to-light decays.
Nevertheless, the previous analyses~\cite{Dem} in $q^{+}\neq 0$ frame
considered only valence contributions neglecting nonvalence
contributions. In this work, we circumvent this problem by calculating
the processes in $q^{+}=0$ frame and analytically continuing to the
timelike region. The $q^{+}=0$ frame is useful because only
valence contributions are needed. However, one needs to calculate
the component of the current other than $J^{+}$ to obtain the form factor
$f_{-}(q^{2})$. Since $J^{-}$ is not free from the zero-mode
contributions even in $q^{+}=0$ frame~\cite{CJ7}, 
we use $J_{\perp}$ instead of $J^{-}$ to obtain $f_{-}$.
In the $q^{+}=0$ frame, we obain the form factors $f_{+}(q^{2})$
and $f_{-}(q^{2})$ using the matrix element of the ``$+$" and
``$\perp$"-components of the current, $J^{\mu}$, respectively, and
then analytically continued to the timelike $q^{2}>0$ region by 
changing $q_{\perp}$ to $iq_{\perp}$ in the form facors.
Our numerical results of the decay rates for
$D$$\to$$\pi(K)$, $D_{s}$$\to$$\eta(\eta')$, and $B\to\pi(D)$ processes
are consistent with the experimental data as summarized in Table 1.
It is interesting to note that our value of $\eta$-$\eta'$ mixing angle,
$\theta_{SU(3)}$=$-19^{\circ}$ presented in Ref.[5], are also in
agreement with the data~\cite{EXP} for $D_{s}$$\to$$\eta(\eta')$ decays. 
More detailed results are given in Ref.[10].
\begin{table}[t]
\caption[]{ Form factors $f_{+}(0)$ and branching ratios (Br.) for various
heavy meson semileptonic decays for $0^{-}$$\to$$0^{-}$.
We use $\theta^{\eta-\eta'}_{SU(3)}$=$-19^{\circ}$ for
$D_{s}\to\eta(\eta')$ decays and the following CKM matrix element:
$|V_{cs}|$=1.04$\pm$0.16,
$|V_{cd}|$=0.224$\pm$0.016,
$|V_{ub}|$=(3.3$\pm$0.4$\pm$0.7)$\times$10$^{-3}$,
and $|V_{bc}|$=0.0395$\pm$0.003 [13].}
\vspace{0.2cm}
\begin{center}
\footnotesize
\begin{tabular}{|c|c|c|c|}
\hline
Processes & $f_{+}(0)$ & Br. & Expt.~\cite{EXP} \\ \hline
$D\to K$  & 0.736 & $(3.75\pm 1.16)\%$ & $(3.66\pm0.18)\%$ \\ \hline
$D\to\pi$ & 0.618 & $(2.36\pm 0.34)\times 10^{-3}$
          & $(3.9^{+2.3}_{-1.1}\pm 0.4)\times 10^{-3}$ \\ \hline
$D_{s}\to\eta$ & 0.421 & $(1.8\pm 0.6)\%$ & $(2.5\pm 0.7)\%$ \\ \hline
$D_{s}\to\eta'$ & 0.585 & $(9.3\pm 2.9)\times 10^{-3}$
          & $(8.8\pm 3.4)\times 10^{-3}$ \\ \hline
$B\to\pi$ & 0.273 & $(1.40\pm 0.34)\times 10^{-4}$ &
$(1.8\pm0.6)\times 10^{-4}$ \\ \hline
$B\to D$  & 0.709 & $(2.28\pm 0.20)\%$ & $(2.00\pm0.25)\% $\\
\hline
\end{tabular}
\end{center}
\end{table}

\section{ Conclusions and Discussions} 
Using the heuristic model presented in Section 2, we discussed the
important issue in the light-front approach, i.e., the calculation
of the timelike form factors. Even though the model was simple, 
it played the role of quidance for the more realistic model calculation.
Since this solvable model provides the full information for the 
nonvalence diagram ( black blob in Fig. 1(b) ), we were able to check
if the analytic continuation of the result in the Drell-Yan-West ($q^{+}=0$) 
frame (without the black blob) indeed reproduces the exact result
generating the peaks analogous to the VMD phenomena.
For more realistic model calculation, we came up with the LFQM~\cite{CJ1}
that described the spacelike form factors quite well. In
this work, we demonstrated that our model can also be applied to the 
timelike exclusive processes, analyzing the semileptonic decays of a
pseudoscalar meson into another pseudoscalar meson with the same 
model in Ref.[5]. 
The form factors $f_{\pm}$ are obtained in $q^{+}=0$ frame and then
analytically continued to the timelike region by changing $q_{\perp}$ to
$iq_{\perp}$ in the form factors. The matrix element of the ``${\perp}$"
component of the current $J^{\mu}$ is used to obtain the form factor
$f_{-}$. Our numerical results are in a good agreement with the available
experimental data. This work broadens the standard light-front frame 
\`{a} la Drell-Yan-West to the timelike form factor calculation. 
The estimation of the zero-mode contribution
has also been presented in Ref.[12]. To the extent that the zero-modes
have a significant contribution to some physical observables~\cite{CJ2},
it seems conceivable that the condensation of zero-modes could lead to
the nontrivial realization of chiral symmetry breaking in the light-front
quantization approach.
The success of our model calculations seems to reveal the
effectiveness of light-front degrees of freedom in exclusive processes.

\section*{Acknowledgments}
This work was supported by the U.S. Department of
Energy(DE-FG-02-96ER4-

\noindent
0947). The North Carolina Supercomputing Center
and the National Energy Research Scientific Computing Center are also
acknowledged for the grant of computing time allocation.
We would like to thank Carl Carlson and Anatoly Radyushkin for organizing
this interesting workshop.


\section*{References}
\begin{thebibliography}{99}
\bibitem{LB} S. D. Drell and T. M. Yan,
\Journal{\PRL}{24}{181}{1970}; G. West, \Journal{\PRL}{24}{1206}{1970};
G. P. Lepage and S. J. Brodsky, \Journal{\PRD}{22}{2157}{1980}.

\bibitem{CJ2} H.-M. Choi and C.-R. Ji, \Journal{\PRD}{59}{034001}{1999}.

\bibitem{CJC} P. L. Chung, F. Coester, and W. N. Polyzou,
\Journal{\PLB}{205}{545}{1988};
W. Jaus, \Journal{\PRD}{44}{2851}{1991};
F. Cardarelli et al., \Journal{\PLB}{332}{1}{1994};
\Journal{\PRD}{53}{6682}{1996}.

\bibitem{CJ4} H.-M. Choi and C.-R. Ji, \Journal{\NPA}{618}{291}{1997};
{\it ibid}. {\bf 56}, 6010 (1997).

\bibitem{CJ1} H.-M. Choi and C.-R. Ji, \Journal{\PRD}{59}{074015}{1999}.

\bibitem{SM}  M. Sawicki and L. Mankiewicz,
\Journal{\PRD}{37}{421}{1988}; L. Mankiewicz and M. Sawicki,
{\it ibid}. {\bf 40}, 3415 (1989).

\bibitem{CJ8} H.-M. Choi and C.-R. Ji, 
``Exploring the timelike region for the elastic form factor in a
scalar field theory", hep-ph/9906225. 


\bibitem{Drell} J. D. Bjorken and S. D. Drell,
{\em Relativistic Quantum Fields} (McGraw-Hill, New York, 1965), pp.
209-282; S. Gasiorowicz, {\em Elementary Particle Physics}
(Wiley, New York, 1966), pp. 348-362.


\bibitem{Isgur} S. Godfrey and N. Isgur, \Journal{\PRD}{32}{189}{1985};
N. Isgur, D. Scora, B. Grinstein, and M. B. Wise,
\Journal{\PRD}{39}{799}{1989};
D. Scora and N. Isgur, \Journal{\PRD}{52}{2783}{1992}.

\bibitem{CJ9}  H.-M. Choi and C.-R. Ji, ``Light-Front Quark Model 
Analysis of Exclusive $0^{-}\to0^{-}$ Semileptonic Heavy Meosn Decays",
to be published in {\em Phys. Lett. B} [hep-ph/9903496]. 

\bibitem{Dem} N.B. Demchuk, I. L. Grach, I. M. Narodetskii, and
S. Simula, \Journal{\PAN}{59}{2152}{1996};
H.-Y. Cheng, C.-Y. Cheng, and C.-W. Hwang,
\Journal{\PRD}{55}{1559}{1997}.

\bibitem{CJ7} H.-M. Choi and C.-R. Ji,
\Journal{\PRD}{58}{071901}{1998}.

\bibitem{EXP} Particle Data Group, C. Caso et al.,
\Journal{\EPJ}{3}{1}{1998}.

\end{thebibliography}
\end{document}